
\magnification=1200
\vsize=22truecm
\hsize=15truecm \tolerance 1000
\parindent=0pt
\baselineskip = 15pt
\lineskip = 1.5pt
\lineskiplimit = 3pt
\font\mybb=msbm10
\def\bb#1{\hbox{\mybb#1}}
\voffset=0pt

\parskip = 1.5ex plus .5ex minus .1ex
{\nopagenumbers
\rightline{KCL-TH-93-14}
\rightline{hep-th/9312010}
\vglue 1truein
\centerline{PARAFERMIONS, $W$ STRINGS AND THEIR BRST CHARGES}
\bigskip
\centerline{Michael Freeman}
\centerline{\&}
\centerline{Peter West}
\bigskip
\centerline{Department of Mathematics}
\centerline{King's College London}
\centerline{Strand, London WC2R 2LS}
\bigskip
\centerline{November 1993}
\vskip .75truein
\centerline{Abstract}
\smallskip
We show how to relate the parafermions that occur in the $W_3$ string
to the standard construction of parafermions.  This result is then
used to show that one of the
screening charges that occurs in parafermionic theories is precisely
the non-trivial part of the $W_3$ string BRST charge.  A way of generalizing
this pattern for a $W_N$ string is explained.
This enables us to construct
the full BRST charge for a $W_{2,N}$ string and to prove the relation
of such a string to the algebra $W_{N-1}$ for arbitrary $N$,
We also show how to calculate part of the BRST charge for a $W_N$ string,
and we explain how our method might be extended to obtain the full BRST
charge for such a string.
\vfil
\eject}


The discovery of extensions of the Virasoro algebra, known as
$W$ algebras, has lead to the study of $W$ string theories based on these
algebras.  In the absence of any clear geometrical understanding of
$W$ algebras, a crucial role is played by the BRST charge in such string
theories.  A review of recent work on $W$ string theory can be
found in [1].

One striking fact that has emerged from a study of the
$W_3$ string is that this theory is closely related to the Ising model.
It was recently shown that the Ising model that arises provides a two-scalar
realization of the $c=1/2$ parafermion theory corresponding to the
coset $SU(2)_2/U(1)$ [2].  A consequence of this fact is that the $W_3$ string
has embedded within it a non-linear $W$ algebra containing an infinite
number of generators $W_s$, one for each spin $s=2, \ldots, \infty$.  In
addition the parafermions could be used to generate the states in the
cohomology of the BRST charge.

The two-scalar realization of the parafermion algebra found in the
$W_3$ string differs from the usual realization of this algebra.  In this
paper, starting from the Wakimoto construction of the $SU(2)$ level $k$
currents [3] and using an alternative bosonization for the bosonic spin 0 and
spin 1 fields used in this construction, we obtain new expressions for the
parafermions.  For the case $k=2$ these coincide with those found in the
$W_3$ string.  For general values of $k$ we show how to relate the two
different realizations of the parafermion algebra, and this enables us to
apply results from the theory of parafermions to the $W_3$ string.  In
particular, we express the known screening charges of the parafermion
theory in terms of the variables of the $W_3$ string.  One of these coincides,
for $k=2$, with the screening charge $S$ that was used in [4] to generate
physical states of the $W_3$ string from three basic physical states and that
was seen to be needed to calculate scattering amplitudes [4].  Another of the
screening charges of the parafermion theory turns out to give the non-trivial
part of the $W_3$ BRST charge.

We argue that this connection between parafermions and $W$ strings
can be generalised, and give evidence that the BRST charge for the
$W_N$ string has a natural decomposition [5,2,6] of the form
$Q(W_N) = Q_0 + Q_1$, where $Q_1$ is just the second of the two
screening charges discussed above, evaluated for $k=N-1$.
We conjecture further that, using the series of cosets
$SU(N-r+1)_r/SU(N-r)_r \times U(1)$, $Q(W_N)$ has a decomposition [5,2,6]
of the form $Q_0^{(r)} + Q_1^{(r)}$,
where now $Q_1^{(r)}$ is a screening charge for the above coset
constructed from the $2(N-r)$ bosonic fields that
are used to describe this coset in the Wakimoto realization [7].

We can also use this procedure to construct an explicit expression for the
BRST charge for the spin-$(2,N)$ string [8] for arbitrary values of $N$.
Using this method it follows immediately that such a string theory will
contain the algebra $W_{N-1}$ [9].

We begin with the Wakimoto realization of the $SU(2)$ level $k$ affine Lie
algebra [3], which is given in terms of a scalar field $\phi$ and a
spin-$(1,0)$ pair of bosonic ghosts $\beta, \gamma$ by
$$
\eqalign{
J^+ &= - \beta \cr
J^3 &= \sqrt{2} \beta\gamma - i \sqrt{k+2} \partial\phi \cr
J^- &= \beta\gamma^2 - i \sqrt{2(k+2)} \partial\phi \gamma +
 k \partial\gamma. }\eqno(1)
$$
The operator product expansions for the fields occurring in these currents
are
$$
\phi(z)\phi(w) = -\log(z-w) \eqno(2)
$$
and
$$
\beta(z)\gamma(w) = - \gamma(z)\beta(w) = {-1 \over z-w}, \eqno(3)
$$
and the energy-momentum tensor is
$$
T = - {1 \over 2} (\partial\phi)^2 - \beta \partial\gamma
 + {i \over \sqrt{2 (k+2)}}\partial^2\phi. \eqno(4)
$$
In order to bosonize the fields $\beta, \gamma$ we first introduce a
spin-(1,0) fermionic ghost pair $(\xi, \zeta)$ and a bosonic field
$\eta$, and write
$$
\beta = \partial\zeta e^{-\eta},\quad \gamma =  \xi e^{\eta}, \eqno(5)
$$
where
$$
\xi(z) \zeta(w) = {1\over z-w}. \eqno(6)
$$
Then $\xi$ and $\zeta$ can then be expressed in terms of a bosonic field
$\chi$
$$
\xi = e^{i \chi},\quad \zeta = e^{- i \chi}, \eqno(7)
$$
with the result that the fields $\beta, \gamma$ can be expressed in terms of
the two bosonic fields $\chi$ and $\eta$ as follows:
$$
\beta = -i \partial\chi e^{-i \chi - \eta},\quad
 \gamma = e^{i \chi + \eta}. \eqno(8)
$$
In terms of $\chi$ and $\eta$, the energy-momentum tensor for the ghosts is
$$
T_{\beta,\gamma} = -{1 \over 2} (\partial\chi)^2 - {1 \over 2}(\partial\eta)^2
 - {i \over 2} \partial^2\chi - {1 \over 2} \partial^2\eta. \eqno(9)
$$

In order to obtain the parafermions $\psi_{\pm 1}$ we divide out by the
$U(1)$ subalgebra generated by $J^3$.  We first observe that, having bosonized
the ghost fields, we can write
$$
J^3 = \sqrt{2}\partial\eta - i \sqrt{k+2} \partial\phi \eqno(10)
$$
so that $J^3$ is the derivative of some field.  This field can then be used
to construct parafermions whose operator product with $J^3$ is regular, namely
$$
\eqalign{
\psi_1 &= - J^+ \exp\left\{-{\sqrt{2} \over k}
\left( \sqrt{2}\eta -  i \sqrt{(k+2)}\phi\right)\right\} \cr
\psi_{-1} &= - k^{-1} J^- \exp\left\{{\sqrt{2} \over k}
\left( \sqrt{2}\eta -  i \sqrt{(k+2)}\phi\right)\right\}
} \eqno(11)
$$
It is convenient to introduce two orthogonal scalar fields $\phi_1$ and
$\phi_2$ by
$$
\eqalign{
\phi_1 &= \chi \cr
\phi_2 &= \sqrt{k+2 \over k} \eta - i \sqrt{2 \over k}\phi
}\eqno(12)
$$
in terms of which the parafermions become
$$
\eqalign{
\psi_1 &= -i \partial\phi_1
\exp\left\{ -i \phi_1 - \sqrt{k+2 \over k}\phi_2\right\} \cr
\psi_{-1} &= -{1 \over k}
\left( i (k+1)\partial\phi_1 + \sqrt{k(k+2)}\partial\phi_2\right)
\exp\left\{ i \phi_1 + \sqrt{k+2 \over k}\phi_2\right\}.
}\eqno(13)
$$
In deriving these expressions we have made use of the formulae
$\beta\gamma = \partial\eta$ and
$\beta\gamma^2 = (i \partial\chi + 2 \partial\eta)e^{i\chi +\eta}$.
The energy-momentum tensor for the fields $\phi_1$ and $\phi_2$ is easily seen
to be
$$
T = -{1 \over 2} (\partial\phi_1)^2 - {1 \over 2} (\partial\phi_2)^2
 - {i \over 2} \partial^2\phi_1
 - {1 \over 2} \sqrt{k \over k+2} \partial^2\phi_2. \eqno(14)
$$

As noted in reference [2], the parafermions that occur in the
$W_3$ string are not of this form.
In fact there is another way to bosonize the ghost fields
$\beta$ and $\gamma$, obtained by taking
$$
\beta = \xi e^{-\eta},\quad \gamma = - \partial\zeta e^{\eta} \eqno(15)
$$
in place of equation (5).  The pair $(\xi, \zeta)$ is bosonized exactly as
before, so that in terms of the scalars $\chi, \eta$ with energy-momentum
tensor given by equation (9) we have
$$
\beta = e^{ i \chi - \eta}, \quad
\gamma = i \partial\chi e^{- i\chi + \eta}. \eqno(16)
$$
We can now follow the same procedure as before to construct parafermions.
In this case the products of $\beta$ and $\gamma$ that we need are
$\beta\gamma = \partial\eta$,
$\gamma^2 = ( i \partial^2\chi - (\partial\chi)^2) e^{-2 i\chi + 2\eta}$ and
$\beta\gamma^2 = (i \partial^2\chi + (\partial\chi)^2 +
2 i\partial\chi \partial\eta)e^{-i\chi+\eta}$.  Using these we obtain the
expressions for the parafermions
$$
\eqalign{
\psi_1 =& \exp\left\{i\phi_1 -\sqrt{k+2 \over k} \phi_2\right\} \cr
\psi_{-1} =& -{1\over k}\left\{ i (k+1)\partial^2\phi_1
+ (k+1)(\partial\phi_1)^2
+ i \sqrt{k(k+2)}\partial\phi_1 \partial\phi_2\right\} \cr
& \times \exp\left\{-i\phi_1 +\sqrt{k+2 \over k} \phi_2\right\}
} \eqno(17)
$$
Here $\phi_1$ and $\phi_2$ have exactly the same forms in terms of
$\chi$, $\eta$ and $\phi$ as previously, in equation (12).

In order to relate these parafermions to those found in the $W_3$ string,
we need to transform $\phi_1, \phi_2$ into the bosonized ghost $\rho$ and
the scalar field $\varphi$ occurring in the $W_3$ string.  For this purpose it
would be sufficient to take $k=2$, but there is some advantage to working
with general $k$ at this stage.  We thus consider fields $\rho$ and
$\varphi$ with energy-momentum tensor
$$
T = - {1 \over 2} (\partial\varphi)^2 - {1 \over 2} (\partial\rho)^2
-{2k +3 \over 2}\sqrt{k \over k+2} \partial^2 \varphi
+ {2k+1 \over 2}i \partial^2\rho . \eqno(18)
$$
The motivation for this expression comes from considering the
$WA_k = W_{k+1}$ string.  This can be built, using the Miura transformation,
from scalar fields $\varphi_2, \ldots, \varphi_{k+1}$ and corresponding ghosts
$b_2, c_2, \ldots, b_{k+1}, c_{k+1}$, and the ghosts can then be bosonized in
terms of fields $\rho_2, \ldots, \rho_{k+1}$ in the usual way.  The
energy-momentum tensor for $\varphi_j, \rho_j$ is given by
$$
T_j = -{1 \over 2} (\partial\varphi_j)^2 -{1 \over 2} (\partial\rho_j)^2
-{2k + 3 \over 2} \sqrt{ (j - 1)j \over (k+1)(k+2)}\partial^2\varphi_j
+ i {(2 j-1) \over 2} \partial^2\rho_j. \eqno(19)
$$
It was observed in [2] that the fields $\varphi_{k+1}, \rho_{k+1}$
together contribute ${2(k-1)/k+2}$ to the central charge, which is
precisely the central charge for the parafermion system discussed above.
The energy-momentum tensor in equation (18) is just that for the
fields $\varphi_{k+1}$ and $\rho_{k+1}$, except that we have written
$\varphi$ and $\rho$ instead of $\varphi_{k+1}$ and $\rho_{k+1}$.

The linear transformation that takes the energy-momentum tensor of
equation (14) into that of equation (18) is
$$
\eqalign{
\phi_1 &= -(k+1) \rho -i \sqrt{k(k+2)}\varphi \cr
\phi_2 &= -i \sqrt{k(k+2)} \rho + (k+1) \varphi.
}\eqno(20)
$$
Setting $k=2$ and substituting the above into equation (17), we indeed
recover the parafermions found in the $W_3$ string.

The transition between the old and new bosonized forms of the ghosts
$\beta, \gamma$ is obtained by making the replacements
$$
-i \partial\chi e^{-i \chi} \rightarrow e^{i \chi}, \quad
 e^{i \chi} \rightarrow i \partial\chi e^{-i \chi}, \eqno(21)
$$
and it is not difficult to see
that this takes the parafermions of equation (13) into those of equation (17).

One advantage of making contact with the well-known formulation of the
parafermions is that one can exploit the knowledge that has been built up
in that formulation.  In particular it is known [10] that there exist three
screening charges $Q_i = \oint j_i\,dz$, where
$$
\eqalign{
j_1 &= e^{i (k+1)\phi_1 + \sqrt{k(k+2)} \phi_2} \cr
j_2 &= e^{i \phi_1} \cr
j_3 &= -i \partial\phi_1 e^{-i \phi_1 - \sqrt{k \over k+2}\phi_2}.
}\eqno(22)
$$
Using the replacement rules given above, the latter two currents become
$$
\eqalign{
j_2 &= i \partial\phi_1 e^{- i \phi_1} \cr
j_3 &= e^{i \phi_1 - \sqrt{k \over k+2}\phi_2}.
}\eqno(23)
$$
While $j_2$ thus becomes a total derivative, so that the corresponding charge
is somewhat trivial, the current $j_3$ can be rewritten in terms of the
fields $\rho, \phi$ as
$$
j_3 = \exp\left( - i \rho + \sqrt{k \over k+2} \phi \right). \eqno(24)
$$
For $k=2$ this is precisely the current for the screening charge $S$ that
was used to relate the different states in the cohomology of the $W_3$
BRST charge.

In order to find the new form of $j_1$ we need the formula
$$
e^{i n \chi} \rightarrow (-1)^{n+1} e^{-i (n+1)\chi}\partial^n e^{i\chi},
\eqno(25)
$$
which we prove by induction.  For $n=1$ this is precisely equation (21).
We now observe that $e^{i (n+1) \chi}$ is the leading order term in the
operator product expansion of $e^{i \chi}$ with $e^{i n \chi}$, so
for the inductive step we
need to consider the operator product of
$-\partial e^{- i \chi} = i \partial\chi e^{-i \chi}$ with
the right hand side of equation (25).  One way to do this is to note that
$$
e^{-i (n+1)\chi}\partial^n e^{i\chi}(w) =
 n! \left[ \oint_w e^{i \chi}, e^{-i(n+1) \chi}(w) \right]. \eqno(26)
$$
It is easy to show that $\oint e^{i \chi}$ commutes with
$-\partial e^{- i \chi}$, so that the operator product we want can be
written as
$$
\eqalign{
-\partial e^{- i\chi}(z)&(-1)^{n+1} e^{-i (n+1)\chi}\partial^n e^{i\chi}(w)\cr
&= (-1)^{n+1} n! \left[ \oint e^{i \chi},-\partial_z e^{- i \chi}(z)
 e^{-i(n+1) \chi}(w)\right].
}\eqno(27)
$$
The required result follows immediately from evaluating the operator
product within the commutator.

Using the above result we find that, in the new formulation of the
parafermions,
$$
j_1 = (-1)^{k+1} e^{i\rho} P_{k+1}(\phi_1), \eqno(28)
$$
where $P_{k+1}$ is a differential polynomial given by
$$
P_{k+1}(\phi_1) = e^{-i\phi_1}\partial^{k+1}e^{i\phi_1}. \eqno(29)
$$
The corresponding charge is then
$$
Q_1 = \oint dz\, e^{i\rho} P_{k+1}(\phi_1). \eqno(30)
$$
We observe that, in the old formulation of the parafermions, the charge
obtained by integrating $j_1$ is manifestly nilpotent, and it follows that
$Q_1$ in equation (30) must also square to zero.
Indeed, for $k=1$ we can evaluate equation (30) to obtain
$$
Q_1 = -6 \oint e^{i\rho} \left( T_\phi + T_\rho \right) \eqno(31)
$$
which is just the bosonized form of the BRST charge for a one-scalar
string with matter energy-momentum tensor $T_\phi$ which has $c=26$.
The fact that the ghost energy-momentum tensor $T_\rho$ does not have
the usual factor of $1/2$ relative to $T_\phi$ is a consequence of working
with bosonized ghosts.  Taking $k=2$ we find that $Q_1$ is given by
$$
\eqalign{
Q_1 = 8\sqrt{2} \oint \Biggl\{ &2 (\partial\phi)^3 +
{21 \over \sqrt{2}} \partial^2\phi\partial\phi +{19 \over 4}\partial^3\phi
- {9i \over 2} \partial^2\rho\partial\phi \cr
&- {9 \over 2} (\partial\rho)^2\partial\phi
- {7 \over 4 \sqrt{2}}\partial^3\rho
-{21 \over 4\sqrt{2}}\partial^2\rho\partial\rho
+ {7i\over 4\sqrt{2}} (\partial\rho)^3 \Biggr\}.
}\eqno(32)
$$
This is just the non-trivial part of the BRST charge for the
$W_3$ string written using the variables introduced in reference [11].

It is clear that the above procedure enables us to calculate a
nilpotent BRST charge for a spin-2 coupled to a spin-{(k+1)} system
for any value of k---the non-trivial part of the BRST charge for such a
system is simply given by equation (30).  Furthermore, it follows from known
results of parafermion systems that this spin-(2,k+1) system will contain
a $W_k$ algebra with $c=2(k-1)/(k+2)$.
The generating fields of this algbera are obtained
as coefficients of powers of $z-w$ arising in the operator product of
$\psi_1$ with $\psi_{-1}$ [12].  Since the parafermions commute with the
charge $Q_1$, so will the $W$ generators, and thus there will be an action of
this $W$ algebra on the physical states of the theory.

It was conjectured in references [2,6] that the BRST charge $Q(W_N)$ for
the Miura representation of $W_N$ can be written in the form
$$
Q(W_N) = Q_0 + Q_1,\eqno(33)
$$
where $Q_1$ is a function of only $\phi_N, b_N$, and $c_N$, and where
$Q_1^2 = 0$.  It is natural to suppose that $Q_1$ is precisely the operator
given in equation (30).  For $W_2$, which is the Virasoro algebra, we have
seen above that $Q_1$ is the usual BRST charge.  For the $W_3$ string the
BRST charge of Thierry-Mieg [13] can be written in the form of equation (33)
by making a suitable field redefinition [11], and we have shown that $Q_1$
is indeed given by equation (30) with $k=2$.  More recently the BRST charge
for $W_4$ has been found [14], and a decomposition of the form (33) has been
found in reference [6].  It has been shown by Duke [15] that in this case also
$Q_1$ is given by equation (30), with $k=3$.

%
%
%
%
It is natural to attempt to generalize this construction so as to obtain
the full BRST charge for the $W_N$ string.  As explained above, the
$W_N$ string can be constructed starting from the Miura realization in terms
of fields $\varphi_2, \ldots,\varphi_N$ and ghosts $\rho_2, \ldots, \rho_N$.
{}From the energy-momentum tensors of these fields one finds [6] that the
fields $\varphi_{r+1}, \ldots,\varphi_N$ and $\rho_{r+1}, \ldots,\rho_N$
give a contribution
$$
c_N^r = (r-1)\left( 1-{r(r+1)\over n(n+1)} \right) \eqno(34)
$$
to the central charge.  This is the central charge of one of the minimal
models of the $W_r$ algebra.  Whereas the usual Miura construction of the
$W_r$ algebra involves $r-1$ scalar fields, the above observation suggests
that there should exist a realization in terms of $2(N-r)$ bosonic fields
for the values of c in equation (34).  It may be possible to obtain such a
realization by generalizing the construction of parafermions as the coset
$SU(2)_k/U(1)$.  In fact, the coset
$$
{ SU(N-r+1)_r \over SU(N-r)_r \times U(1)}\eqno(35)
$$
can be seen to have the central charge $c_N^r$, by using the formula
$c=k\, {\rm dim}G/(k+h)$ for the central charge of a level-$k$ affine Lie
algebra $G$ with Coxeter number $h$.  Furthermore, the Wakimoto construction
of $SU(2)_k$ has been generalized to $SU(n)_k$ giving a realization of
$SU(n)_k$ in terms of dim$SU(n)$ bosonic fields [7].  Consequently, the
generalized parafermions corresponding to the coset of equation (35) would
have a description in terms of $2(N-r)$ bosonic fields.  The form of these
parafermions is unknown, but we note that this coset has the same
$W$ algebra corresponding to it as
$SU(r)_{N-r}\times SU(r)_1/SU(r)_{N-r+1}$ [16], which has a description in
terms of $r-1$ bosonic fields and provides the standard unitary representations
of $W_r$ [17].

We therefore propose that it should be possible to construct some
generalized parafermions from the coset of equation (35) which can be
expressed in terms of the fields
$\varphi_{r+1}, \ldots,\varphi_N$ and $\rho_{r+1}, \ldots, \rho_N$.  These
parafermions should generate $W_r$, and they should possess a screening
charge $Q_1^{(r)}$ such that the BRST charge for the $W_N$ string takes
the form
$$
Q(W_N) = Q_0^{(r)} + Q_1^{(r)}\eqno(36)
$$
with $(Q_1^{(r)})^2 = 0$.  Clearly the case $r=N-1$ reduces to
$SU(2)_{N-1}/U(1)$ which was discussed above, and for $r=2$ we should
recover the full screening charge of the Miura realization of the
$W_N$ string.  In this way we would be able to obtain for the quantum
case a series of nested BRST charges, analogous to the classical BRST
charges obtained in reference [6].
Further details will be given elsewhere [18].

We have seen how the non-trivial part of the BRST charge for the
$W_3$ string and for the $W_{2,s}$ string can be constructed using
parafermions.  It is of interest also to consider the action of the
parafermions on the physical states of the $W_3$ string, contained in
the cohomology of its BRST charge.  These states can be found by acting
with screening charges $S$, $S_1$ and $S_2$ and with the picture-changing
operator $P$ on three basic states [4].  In the notation of [4], the three
basic states are $V(a,0) = \phi(h,0) V_X(a)$, where $a=1,15/16, 1/2$, $h=1-a$,
$\phi(h,0)$ is an Ising primary field of weight $h$, and $V_X(a)$ is a vertex
operator constructed from the fields of the matter sector.  We can then
construct the following series of states: $V(15/16,m) = (S^2P)^m V(15/16,0)$,
and $V(a,m) = (S^3P)\bar V(a,m-1), \bar V(a,m) = SPV(a,m)$ for intercepts
$a=1$ and $a=1/2$ [4].  We can also find series of conjugate states.
Thus $V(1,n)$, $\bar V(1,n)$, $V(1/2,n)$ and $\bar V(1/2,n)$ have
conjugates $P(S_1P)^n \bar V(1,0)$, $P(S_1P)^n V(1,0)$,
$P(S_1P)^{n+1}\bar V(1/2,0)$ and $P(S_1P)^{n+1} V(1/2,0)$ respectively,
while $V(15/16,n)$ has conjugate vertices $P(S_1P)^{n/2} V(15/16,0)$ for
$n$ even and $P(S_1P)^{(n+1)/2} V(15/16,1)$ for $n$ odd.  It is straightforward
to verify that these conjugates indeed have the correct $\phi$ and $\rho$
momenta to give a non-zero scalar product.  The low level conjugates were
found in collaboration with B. Nilsson [19].

The $W_3$ string also includes the discrete physical state [20]
$$
D(0) = (c e +{5 \over 87}\sqrt{58} i \partial e \, e) e^{i({8 i Q/7})}.
\eqno(37)
$$
These
vertices have the same $\phi$ momenta as $V(1,n)$, and we can create
the following infinite number of discrete physical vertices
$$
S^3P\bar D(n-1) = D(n), \quad \bar D(n) = (SP) D(n) \eqno(38)
$$
and similarly form their conjugates.

As outlined in reference [2], physical states can also be created using the
parafermions $\psi_{\pm 1}$.  The parafermion fields $\psi_{\pm 1}$ are
given by $\psi_1 = \phi(1/2,1)$ and $\psi_{-1}= \phi(1/2,2)$, and, using the
notation of reference [21], the parafermionic primary fields
$\phi^0_0$, $\phi^1_1$ and $\phi^1_{-1}$ are given by $\phi(1,1)$,
$\phi(1/16,1)$ and $\phi(1/16,2)$ respectively.  In fact, one can act
on the basic state $\phi(a,0)$ with $\psi_{-1}$ to obtain the states above,
and with $\psi_1$ to obtain their conjugates.  Let us first
consider Ising weight $1/16$ states.  $\psi_{-1}(z) \phi(1/16,n)$ has an
operator product expansion in powers of $(z-w)^{p + 1/2}$, $p \in \bb{Z}$,
all the coefficients  of which are annihilated by $Q_1$.
For small enough $n$ the most singular term in the OPE is
$(z-w)^{-1/2}$, and the coefficient of this term is a primary field of
weight $1/16$ which is $\phi(1/16,n+1)$.  For larger $n$ it is no longer the
case that $(z-w)^{-1/2}$ is the most singular term, but one can construct
$\phi(1/16, n+1)$ out of the coefficient of $(z-w)^{-1/2}$ by subtracting
Virasoro descendents of the coefficients of $(z-w)^{-r}$ for $r \ge 1/2$.
A similar construction works for $\psi_1$ acting on $P\phi(1/16,0)$ to
create the conjugates.

In a similar way we can obtain $\phi(1/2,n)$ from the operator product
expansion $\psi_{-1}(z) \phi(0,n)(w)$, and $\phi(0,n+1)$ from
$\psi_{-1}(z) \phi(1/2,n)(w)$.  We can also get $\phi(a,n)$ for
$a=1/2,0$ in this way, and acting with $\psi_1$ gives the conjugate states.
Further details of this construction will be given elsewhere.

Acknowledgement

M. Freeman is grateful to the SERC for financial support.
\vfil
\eject

{{\bf References}}
\parskip 0pt
\item{[1]} P. West, ``A review of $W$ strings,'' preprint
G\"oteborg-ITP-93-40
\item{[2]} M. Freeman and P. West, Phys. Lett. B314 (1993) 320.
\item{[3]} M. Wakimoto, Comm. Math. Phys. 104 (1986) 605;
A. B. Zamolodchikov, unpublished.
\item{[4]} M. Freeman and P. West, Int. J. Mod. Phys. A8 (1993) 4261.
\item{[5]} E. Bergshoeff, H. J. Boonstra, M. de Roo, S. Panda and
A. Sevrin, ``On the BRST operator of W-strings,'' preprint UG-2/93.
\item{[6]} E. Bergshoeff, H. J. Boonstra, S. Panda and M. de Roo,
``A BRST analysis of $W$-symmetries,'' preprint UG-4/93.
\item{[7]} B. L. Feigin and E. V. Frenkel, Comm. Math. Phys. 128 (1990) 161;
K. Ito and Y. Kazama, Mod. Phys. Lett. A5 (1990) 215;
M. Kuwahara and H. Suzuki, Phys. Lett. B235 (1990) 52.
\item{[8]} H. Lu, C. N. Pope and X. J. Wang, ``On higher-spin
generalisations of string theory,'' preprint CTP TAMU-22/93, hep-th/9304115.
\item{[9]} H. Lu, C. N. Pope, X. J. Wang and K. Thielemans,
``Higher-spin strings and $W$ minimal models,''
preprint CTP TAMU-43/93, hep-th/9308114.
\item{[10]} P. Bouwknegt, J. McCarthy and K. Pilch, Comm. Math. Phys. 131
(1990) 125; T. Jayaraman, K. S. Narain and M. H. Sarmadi, Nucl. Phys. B343
(1990) 418.
\item{[11]} H. Lu, C. N. Pope, S. Schrans and X-J. Wang, Nucl. Phys. B408
(1993) 3.
\item{[12]} F. J. Narganes-Quijano, Ann. Phys. (NY) 206 (1991) 494;
Int. J. Mod. Phys. A6 (1991) 2635.
\item{[13]} J. Thierry-Mieg, Phys. Lett. B197 (1987) 368.
\item{[14]} K. Hornfeck, ``Explicit construction of the BRST charge for
$W_4$,'' preprint DFTT-25/93, hep-th/9306019; Chuang-Jie Zhu, ``The BRST
quantization of the non-linear $WB_2$ and $W_4$ algebras,''
preprint SISSA/77/93/EP.
\item{[15]} G. Duke, unpublished.
\item{[16]} P. Bowcock and P. Goddard, Nucl. Phys. B305 [FS23] (1988) 685;
D. Altschuler, Nucl. Phys. B313 (1989) 293.
\item{[17]} M. Kuwahara and H. Suzuki, Phys. Lett. B235 (1990) 52.
\item{[18]} G. Duke, M. Freeman and P. West, work in progress.
\item{[19]} B. Nilsson and P. West, unpublished.
\item{[20]} C. N. Pope, E. Sezgin, K. S. Stelle and X. J. Wang,
Phys. Lett. B299 (1993) 247.
\item{[21]} A. B. Zamolodchikov and V. A. Fateev, Sov. Phys. JETP {\bf 62}
1985 215.

\end